# Advancing Biomedical Signal Security: Real-Time ECG Monitoring with Chaotic Encryption

**Beyazıt Bestami YÜKSEL, Ayşe YILMAZER METİN**

Istanbul Technical University

*yukselbe18@itu.edu.tr, yilmazerayse@itu.edu.tr*

**Abstract**

The real-time analysis and secure transmission of electrocardiogram (ECG) signals are critical for ensuring both effective medical diagnosis and patient data privacy. In this study, we developed a real-time ECG monitoring system that integrates chaotic encryption to protect the integrity and confidentiality of ECG signals during acquisition, transmission, and storage. By leveraging the logistic map as the chaotic function for encryption, our system offers a highly secure framework that dynamically encrypts ECG signals without adding significant latency. To validate the system's reliability, we applied a series of security tests. The results demonstrate that chaotic encryption is effective in enhancing data security, as evidenced by high entropy values and strong key sensitivity, ensuring protection against common cryptographic attacks. Additionally, the system's real-time disease detection model, based on deep learning, operates seamlessly with encrypted data, providing accurate diagnosis without compromising security. Our findings indicate that chaotic encryption, paired with real-time analysis, is a powerful method for protecting sensitive medical data, making this approach particularly relevant for telemedicine and remote patient monitoring applications. The success of this system highlights its potential for broader application to other biomedical signals, providing a secure infrastructure for the future of digital health.

## Introduction

In recent years, the growing use of medical devices for continuous patient monitoring has highlighted the critical need for securing sensitive biomedical data, such as electrocardiograms (ECG). Ensuring the privacy and integrity of patient information is paramount, especially given the risks of unauthorized access. Traditional encryption methods, while effective, often introduce computational overhead that can hinder real-time systems where immediate data processing is indispensable. Chaotic encryption techniques have emerged as a promising alternative, offering both high security and efficiency due to their unpredictability and sensitivity to initial conditions. This study presents a real-time ECG monitoring system that incorporates chaotic encryption to safeguard biomedical signals during transmission and storage. The system's primary goal is to ensure secure data handling while providing real-time disease diagnosis. By employing the logistic map for chaotic encryption, the system scrambles ECG signals, making them difficult to interpret without the correct decryption key, thus ensuring data confidentiality. In addition to its security features, the system integrates a machine learning model trained to detect abnormal heart rhythms and other cardiovascular conditions in real-time. This combination of secure monitoring and automated disease detection enhances the system's practical utility, enabling healthcare professionals to quickly respond to potential issues. The real-time functionality ensures that medical professionals have access to immediate insights, improving patient care and timely intervention. The key advantage of this system lies in its dual functionality: not only does it securely transmit and store ECG data, but it also performs real-time analysis for disease diagnosis. This makes it highly suitable for applications such as remote patient monitoring and telemedicine, where both data security and rapid medical response are essential. The integration of chaotic encryption ensures that security is maintained without compromising performance, offering a comprehensive and efficient solution for modern healthcare needs.

**Related Works**

Numerous studies have applied chaotic encryption techniques to secure biomedical signals, particularly ECG, in telemedicine applications. Murillo-Escobar et al. [1] proposed a chaos-based cryptographic algorithm using the Badola map for real-time ECG transmission in embedded systems, demonstrating robustness against cryptographic attacks. Pandian and Ray [2] introduced a dynamic hash key-based stream cipher for real-time ECG transmissions, implemented on FPGA hardware with improved data throughput. Sufi and Khalil [3] developed a chaos-based encryption technique for time-critical ECG packets, ensuring security through correlation and entropy analyses. Ghubaish et al. [4] applied chaotic encryption to multimedia data in IoT-based surveillance, emphasizing its efficiency for large data sets in healthcare systems. Algarni et al. [5] presented an encryption system using ECG signals and masking signals, offering enhanced security in noisy environments. Rajasree and Kumar [6] used Henon and Baker chaotic maps for the secure transmission of ECG signals, proving resilience against cryptographic attacks. Mekki et al. [7] proposed chaotic encryption for multimedia data in IoT healthcare surveillance, highlighting its advantages over traditional methods. Ye and Huang [8] developed an image encryption algorithm based on autoblocking and ECG signals, showing high resistance to differential attacks. Bai et al. [9] introduced a hybrid cryptosystem using chaotic maps and biometric authentication to secure ECG signals in Body Area Networks. Ramírez-Villalobos et al. [10] explored a multi-scroll chaotic system for encrypting ECG signals in wireless networks, ensuring data integrity through master-slave synchronization. Yavuz used a novel parallel processing architecture for accelerating image encryption based on chaos [11]. The security tests used by the author also inspired this study.

Our proposed work differs from previous studies by integrating chaotic encryption with real-time disease detection based on deep learning models, offering a dual approach to secure and analyze ECG signals simultaneously. While prior studies focus primarily on the encryption of ECG signals, this work extends functionality by incorporating a CNN-based disease detection model trained on the MIT-BIH Arrhythmia Database. Additionally, the implementation of chaotic encryption using logistic maps in real-time is tailored for embedded systems, enhancing both security and computational efficiency. Our study also applies a comprehensive set of security tests, including NIST frequency analysis, Shannon entropy, avalanche effect, key sensitivity, and correlation tests, to evaluate the robustness of the encryption scheme.

**Methodology and Proposed System**

The system for real-time ECG signal monitoring and disease diagnosis is based on a three-lead ECG preamplifier, connected to a computer via a serial port. Real-time data reading operations from the serial port to be run on different platforms have been performed by the author of this article in his previous works [12][14]. The ECG signals are sampled at a frequency of 500 Hz, a standard rate in biomedical signal processing to capture the necessary detail for accurate signal interpretation. Each byte of data from the preamplifier is received in real time, and processed in 300-sample segments. Each segment corresponds to approximately 0.6 seconds of ECG signal. This continuous stream of ECG data is then fed into both a machine learning model for disease diagnosis and a chaotic encryption system for securing the transmission of biomedical data. The algorithm of the implemented application is given below. The ECG preamplifier used is completely portable with 3 probes connected to the body.

```
Algorithm: Real-Time ECG Signal Processing and Disease Detection with Chaotic Encryption
1: Initialize serial port for real-time ECG data acquisition
2: Load pre-trained CNN model for disease detection
3: Set logistic map parameters for chaotic encryption (key = 0.5, r = 3.99)
4:
5: while True do
6:    if data available on serial port then
7:       Read ECG data segment
8:       Center the ECG signal (subtract 128 from each sample)
9:
10:      // Step 1: Chaotic Encryption
11:      for each sample in ECG data do
12:         encrypted_sample = logistic_map_encrypt(sample, key, r)
13:         Append encrypted_sample to encrypted_data list
14:      end for
15:
16:      // Step 2: Chaotic Decryption
17:      for each encrypted_sample in encrypted_data do
18:         decrypted_sample = logistic_map_decrypt(encrypted_sample, key, r)
19:         Append decrypted_sample to decrypted_data list
20:      end for
21:
22:      // Step 3: Preprocess and Normalize Data
23:      Preprocess decrypted_data (normalize: subtract mean, divide by std deviation)
24:      Reshape data to match model input shape
25:
26:      // Step 4: Disease Prediction
27:      predicted_class = CNN_Model.predict(decrypted_data)
28:      Display predicted_class on GUI
30:      // Step 5: Display Results
31:      Update real-time plot with decrypted_data (raw and filtered)
32:      Update GUI with predicted class and processing latency
34:   end if
35:   Sleep for 0.01 seconds (to reduce CPU usage)
36: end while
// Logistic Map Encryption Function
Function logistic_map_encrypt(sample, key, r)
1: x = key
2: for i = 1 to length(sample) do
3:    x = r * x * (1 - x)
4:    encrypted_sample = sample ⊕ (int(x * 255))  // XOR operation
5: return encrypted_sample
End Function
// Logistic Map Decryption Function
Function logistic_map_decrypt(encrypted_sample, key, r)
1: x = key
2: for i = 1 to length(encrypted_sample) do
3:    x = r * x * (1 - x)
4:    decrypted_sample = encrypted_sample ⊕ (int(x * 255))  // XOR operation
5: return decrypted_sample
End Function
End Algorithm
```

**Signal Acquisition:** The sampled ECG signal at time t, denoted as s(t), is described by:

**s(t)=V$_{ECG}$(t)+N(t),** where $V_{ECG}(t)$ is the voltage of the ECG signal at time t, and N(t) represents noise that may come from external sources (e.g., muscle noise, electrode motion, etc.). The data is collected in 8-bit integers ranging from −128 to +127, and then normalized for further processing.

**Chaotic Encryption:** In this system, a logistic map is employed as the core of the chaotic encryption mechanism. The logistic map is a mathematical function that generates a chaotic sequence based on the following recursive relationship:

**x$_n$+1 =r · x$_n$ · (1− x$_n$),** where **r** is a control parameter (with a typical value r=3.99), and **x$_n$** is the state at iteration n. The logistic map produces a sequence of values that appear random but are deterministically generated, making it a suitable candidate for encryption. This chaotic sequence is used to encrypt the ECG signal through an XOR operation.

**Encryption Process:** The encryption process for each ECG signal s(t) at time t can be expressed as:

**E(s(t)) = s(t) ⊕ [xn ·255],** where E(s(t)) is the encrypted value of the ECG signal, and [ $x_n$ · 255] represents the chaotic sequence value scaled to an 8-bit integer range. The XOR operation ensures that the original ECG signal is masked by the chaotic sequence, making the data secure.

**Decryption Process:** Since XOR is a reversible operation, the decryption process is straightforward:

**D(E(s(t))) = E(s(t)) ⊕ [x_n * 255] = s(t),** where *D(E(s(t)))* is the decrypted signal, which matches the original ECG signal *s(t)* when the same chaotic sequence is applied.

**Disease Diagnosis Model:** The real-time ECG monitoring system integrates a machine learning model to diagnose heart abnormalities based on the collected ECG data. The diagnosis model is a convolutional neural network (CNN) trained on the MIT-BIH Arrhythmia Database. The model is designed to detect different types of arrhythmias based on R-peak detection and feature extraction. R-peak detection is carried out using the GQRS algorithm, which identifies the QRS complex in the ECG signal. Once the R-peaks are detected, segments of the ECG signal around each R-peak are extracted for classification.

**R(t) = max(V_ECG(t)),** where R(t) represents the R-peak. For each detected R-peak, a segment of N=180 samples (0.36 seconds) is extracted, providing a window around the R-peak that is used as input to the CNN model.

**Convolutional Neural Network (CNN) Architecture:** The CNN used for real-time disease diagnosis consists of the following layers:

1. **Conv1D Layer**: Extracts local features from the ECG signal with 32 filters and kernel size of 3.
2. **Dropout Layer**: Reduces overfitting by randomly dropping neurons with a probability of 50%.
3. **Flatten Layer**: Transforms the multi-dimensional data into a single dimension.
4. **Dense Layer**: Fully connected layer with 100 neurons and ReLU activation.
5. **Output Layer**: A softmax layer with 5 output classes corresponding to different arrhythmias: Normal (N), Left Bundle Branch Block (LBBB), Right Bundle Branch Block (RBBB), Atrial Premature Contraction (APC), and Ventricular Premature Contraction (VPC).

The loss function used during training is categorical crossentropy, which is suitable for multi-class classification, and the optimizer is Adam, which is known for its efficiency in handling large datasets.

$$Loss = -\sum_{i=1}^{N} y_i . log(\hat{y}_i)$$

where $y_i$ is the true label and $\hat{y}_i$ is the predicted probability for class i. The CNN model is trained on segments of ECG data and then deployed in real-time to predict the arrhythmia class for incoming data segments.

## Security Analysis

In order to validate the robustness of the encryption scheme, several security tests were applied to the encrypted ECG data:

*1. NIST SP800-22 Frequency Test*

This test checks the randomness of the encrypted data by evaluating whether the number of '0's and '1's in the binary representation of the signal is approximately equal. A passing test indicates that the encryption process has produced data that exhibits randomness akin to truly random data. In this study, NIST SP800–22 statistical test suite [13] was used to test the randomness of output sequence generated by the proposed cryptosystem.

*2. Shannon Entropy*

Shannon entropy is used to quantify the randomness of the encrypted signal. It is defined as:

$$H(x) = -\sum_{i=1}^{n} P(x_i) log_2 P(x_i)$$

where $P(x_i)$ is the probability of the i-th symbol in the encrypted sequence. Higher entropy values indicate higher randomness.

*3. Avalanche Effect*

The avalanche effect measures how a small change in the input (e.g., one bit) leads to a significant change in the encrypted output. It is quantified as the ratio of changed bits in the output to the total number of bits:

$$A = \frac{\text{Number of changed bits}}{\text{Total number of bits}}$$

A value close to 1 indicates that the encryption algorithm exhibits strong sensitivity to small changes, which is a desirable property in encryption schemes.

*4. Key Sensitivity Test*

Key sensitivity measures the degree to which small changes in the encryption key affect the output. Ideally, a small perturbation in the key (e.g., changing the key by 0.01) should cause the encrypted output to be completely different, thereby ensuring security.

*5. Correlation Test*

The correlation test measures the linear relationship between the original and encrypted signals. A correlation close to 0 implies that the encrypted signal shares no linear resemblance to the original signal, ensuring that the encryption process has sufficiently obfuscated the data.

$$\rho(X, Y) = \frac{\text{Cov}(X, Y)}{\sigma x \sigma y}$$

where Cov(X,Y) is the covariance of the original and encrypted data, and σX and σY are their standard deviations, respectively.

# Experimental Results

This section presents the results obtained from the chaotic encryption and decryption process applied to real-time ECG signals. The results indicate that the chaotic encryption method based on the logistic map is highly effective for securing real-time ECG signals. The NIST Frequency Test and Shannon Entropy results demonstrate that the encrypted signals exhibit strong randomness, making it difficult for potential attackers to detect any patterns or correlations in the data. The Avalanche Effect and Key Sensitivity results confirm that the system provides high diffusion and sensitivity to small changes, ensuring that even minor modifications in the input or key lead to significantly different encrypted outputs. The low correlation between the encrypted and decrypted signals further supports the robustness of the encryption system, as it ensures that the encrypted data does not reveal any information about the original signals. All the test results shown in Table 1. Additionally, the histograms illustrate that while the encrypted data is uniformly distributed, the decrypted signals faithfully represent the original ECG signals. Overall, the chaotic encryption method applied in this study successfully ensures both the confidentiality and integrity of real-time ECG signals. The system is capable of real-time disease diagnosis while maintaining a high level of security, making it well-suited for use in secure telemedicine and remote patient monitoring applications. These findings demonstrate the potential of chaotic encryption methods for enhancing biomedical signal security, particularly in environments where privacy and real-time processing are of paramount importance.

The system as shown in Figure 1 provides an overview of the system and Figure 2 demonstrates real-time plots of the raw ECG signal, the encrypted ECG signal, and the decrypted ECG signal. This allows the user to observe the signal in real-time as it is processed by the system. The interface also displays the disease prediction results, including the processing latency for each prediction.

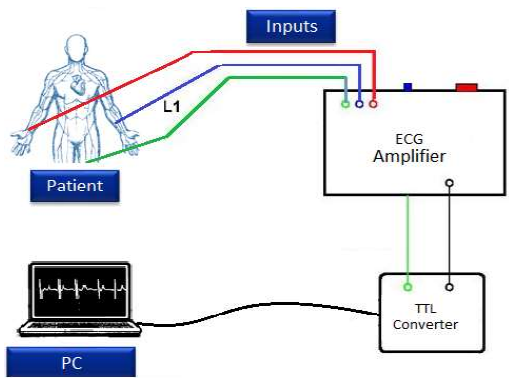

Figure 1: Hardware Setup of System

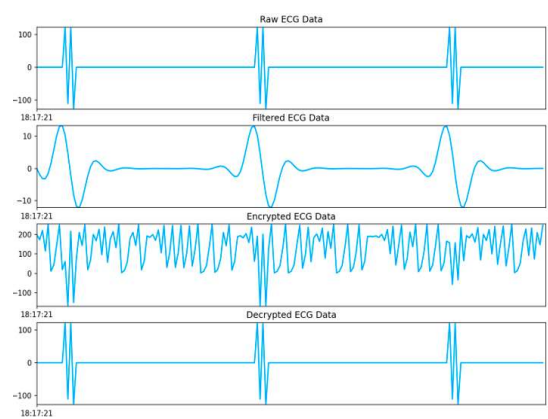

Figure 2: Demonstration Results: Synchronized Raw, Filtered, Encrypted and Decrypted Data Plots with Timestamp

**Histograms of Encrypted and Decrypted Data:** The histograms of the encrypted and decrypted ECG signals provide a visual representation of the data distribution. The histogram of the encrypted data shows a uniform distribution (Figure 2 - a), which is characteristic of a well-encrypted signal, indicating a lack of discernible patterns or biases. Conversely, the histogram of the decrypted data (Figure 2 - b) closely resembles that of the original ECG signal, demonstrating that the decryption process accurately recovers the original signal without significant distortion.

| Test | Result |
| --- | --- |
| NIST Frequency | True |
| Shannon Entropy | 7.935 |
| Avalanche Effect | 1.0 |
| Key Sensitivity | 0.9917 |
| Correlation | 0.0075 |

Table 1: Security Tests Results

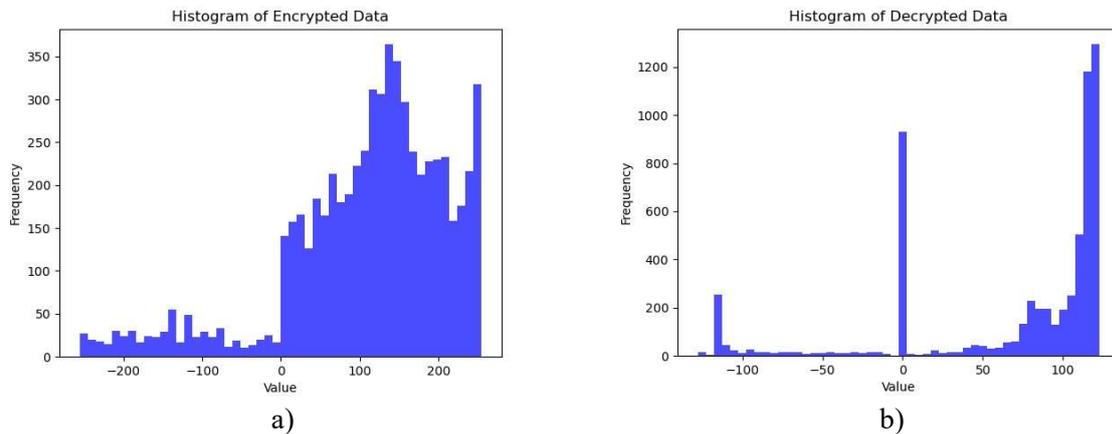

a)　　　　　　　　　　　　　　　　　　b)

Figure 3: Histogram Analysis: (a) Encrypted Data and Histograms Plots; (b) Related Decrypted Data and Histogram Plots

**Conclusion & Future Works**

This study demonstrates the effectiveness of chaotic encryption methods in ensuring the secure transmission and storage of real-time ECG signals. By leveraging nonlinear dynamic systems such as the logistic map, chaotic encryption provides a high level of security. The results obtained in this study validate the success of chaotic encryption in safeguarding ECG signal confidentiality. The security tests applied have shown strong performance in terms of randomness, diffusion, key sensitivity, and signal integrity. These findings confirm that chaotic encryption is highly effective in securing ECG signals. The potential of this technology in real-time medical data security is significant, particularly for telemedicine and remote patient monitoring systems where the confidentiality of health data is paramount. As shown in this study, chaotic encryption methods can be applied not only to ECG signals but also to other biomedical signals, ensuring their security. In the future, these methods should be further tested and refined for broader applications in biomedical fields

Future research can explore how chaotic encryption algorithms can be adapted and optimized for other medical signals, such as EEG (Electroencephalography) or EMG (Electromyography). Each signal type has its unique characteristics and requirements for encryption, and tailoring chaotic systems to these signals could expand the scope of secure medical data transmission. To enhance the real-time encryption performance, hardware-based accelerators such as CNN (Convolutional Neural Network) accelerators or FPGA (Field Programmable Gate Arrays) could be integrated. These technologies can provide faster encryption processes, making chaotic encryption more viable for real-time monitoring, even in high-volume signal environments. As telemedicine and remote patient monitoring systems continue to grow, integrating cloud-based solutions for secure storage and signal analysis is a promising direction. Future work could involve designing secure cloud architectures where encrypted biomedical data can be safely stored, processed, and analyzed in real-time, allowing for both enhanced security and accessibility.

# References

[1] D. Murillo-Escobar, C. Cruz-Hernández, R.M. López-Gutiérrez, M.A. Murillo-Escobar, "Chaotic encryption of real-time ECG signal in embedded system for secure telemedicine," *Integration, the VLSI Journal*, vol. 89, pp. 261–270, 2023.

[2] K. K. Soundra Pandian, K. C. Ray, "Dynamic Hash key-based stream cipher for secure transmission of real time ECG signal," *Security and Communication Networks*, vol. 9, pp. 4391–4402, 2016.

[3] M.H. Sufi, I. Khalil, "A chaos-based encryption technique to protect ECG packets for time-critical telecardiology applications," *Security and Communication Networks*, vol. 4, pp. 1615-1628, 2011.

[4] M.A. Ghubaish, A. AlShamsi, A. AlShehhi, N.H. AlMansoori, "A real-time chaotic encryption for multimedia data and application to secure surveillance framework for IoT system," Journal of Systems Architecture, vol. 130, pp. 101973, 2023.

[5] D. Algarni, N. F. Soliman, H. A. Abdallah, F. E. Abd El-Samie, "Encryption of ECG signals for telemedicine applications," Multimedia Tools and Applications, vol. 80, pp. 10679–10703, 2021.

[6] G. Rajasree, R. Mathusoothana S. Kumar, "Secure transmission and monitoring of ECG signals based on chaotic mapping algorithms," Automatika, vol. 65, no. 3, pp. 957-972, 2024.

[7] N. Mekki, M. Hamdi, T. Aguili, T. Kim, "A Real-Time Chaotic Encryption for Multimedia Data and Application to Secure Surveillance Framework for IoT System," IEEE Transactions on Multimedia, 2018.

[8] G. Ye, X. Huang, "An Image Encryption Algorithm Based on Autoblocking and Electrocardiography," IEEE Multimedia, 2016.

[9] F. B. Bai, "Encryption Algorithm for Biomedical Signals Using Chaotic Systems," Journal of Wireless Personal Communications, vol. 122, pp. 987-1003, 2020.

[10] J. R. Cárdenas-Valdez, R. Ramírez-Villalobos, C. Ramirez-Ubieta, E. Inzunza-Gonzalez, "Enhancing Security of Telemedicine Data: A Multi-Scroll Chaotic System for ECG Signal Encryption and RF Transmission," Entropy, vol. 26, no. 787, 2024.

[11] Yavuz, E. (2021). A new parallel processing architecture for accelerating image encryption based on chaos. Journal of Information Security and Applications, 63, 103056.

[12] Yuksel, Beyazit Bestami, Yazici Pasa, & Bilgin Gokhan, "Gerçek Zamanlı EKG İşaretlerinin Mobil Sistemde İzlenmesi", 2. Ulusal Biyomedikal Cihaz Tasarımı ve Üretimi Sempozyumu, 16 Mayıs 2017.

[13] Bassham III LE. SP 800-22 rev. 1a, a statistical test suite for random and pseudorandom number generators for cryptographic applications. National Institute of Standards & Technology; 2010. Technical report sp800-22.

[14] Yuksel, Beyazit Bestami. EKG işaretlerinin gömülü sistem ile izlenmesi. 2011. Master's Thesis. Marmara Universitesi (Turkey).